\documentclass[%
 reprint,
 amsmath,amssymb,
 aps,
]{revtex4-2}

\usepackage{graphicx}
\usepackage{dcolumn}
\usepackage{bm}

\usepackage{hyperref}
\usepackage{color}
\usepackage{physics}
\usepackage{amsmath}

\DeclareMathOperator*{\argmin}{arg\,min}

\begin{document}

\preprint{APS/123-QED}

\title{Predicting rate kernels via dynamic mode decomposition}

\author{Wei Liu}
\affiliation{Department of Chemistry, School of Science, Westlake University, Hangzhou 310024 Zhejiang, China}
\affiliation{Institute of Natural Sciences, Westlake Institute for Advanced Study, Hangzhou 310024 Zhejiang, China}
\author{Zi-Hao Chen}%
\affiliation{Department of Chemical Physics, University of Science and Technology of China, Hefei, Anhui 230026, China
}%

\author{Yu Su}%
\affiliation{Department of Chemical Physics, University of Science and Technology of China, Hefei, Anhui 230026, China
}%


\author{Yao Wang}
\email{wy2010@ustc.edu.cn}
\affiliation{Department of Chemical Physics, University of Science and Technology of China, Hefei, Anhui 230026, China
}%

\author{Wenjie Dou}
\email{douwenjie@westlake.edu.cn}
\affiliation{Department of Chemistry, School of Science, Westlake University, Hangzhou 310024 Zhejiang, China}
\affiliation{Institute of Natural Sciences, Westlake Institute for Advanced Study, Hangzhou 310024 Zhejiang, China}
\affiliation{Department of Physics, School of Science, Westlake University, Hangzhou 310024, Zhejiang, China}


\date{\today}

\begin{abstract}
Simulating dynamics of open quantum systems is sometimes a significant challenge, despite the availability of 
various exact or approximate methods.
Particularly when dealing with complex systems, 
the huge computational cost will largely limit the applicability of these methods.
We investigate the usage of dynamic mode decomposition (DMD) to evaluate the rate kernels in quantum rate processes.
DMD is a data-driven model reduction technique that characterizes the rate kernels using snapshots collected from
a small time window, allowing us to  predict the long-term behaviors with only a limited number of samples.
Our investigations show that whether the external field is involved or not, the DMD can
give accurate prediction of the result compared with the traditional propagations, and simultaneously 
reduce the required computational cost.

\end{abstract}

\maketitle


\section{Introduction}
Dynamic mode decomposition (DMD) was introduced by Schmid \cite{schmid2010dynamic} in fluid dynamics to study spatio-temporal coherent structures from high-dimensional data. 
Built upon the proper orthogonal decomposition (POD) and the singular value decomposition (SVD), DMD method aims to efficiently reduce the dimensionality of complex systems. 
Different from the POD or SVD which disregards the temporal information,
DMD offers a modal decomposition method which not only achieves the dimensionality reduction  but also produces the dynamical behaviors of these modes.
Shortly after the initial development of the DMD algorithm \cite{schmid2010dynamic}, Rowley, Mezic, and their collaborators established the connection between the DMD and the Koopman's theory \cite{rowley2009spectral}. %
Seeking to identify the most suitable temporal frequencies and spatial modes \cite{tu2013dynamic}, DMD essentially serves as an approximation to the Koopman's operator which governs the dynamics of a high-dimensional system. 
Using
the linear DMD framework to study nonlinear dynamical systems has attract a lot of research interests  \cite{kutz2016dynamic}.
Recently, DMD has emerged as a versatile algorithm for the data-driven characterization of high-dimensional systems. This algorithm is applicable to both experimental and numerical data, 
and combines the advantageous features of the SVD for spatial dimensionality reduction and the Fast Fourier Transform (FFT) for identifying temporal frequencies \cite{chen2012variants,kutz2016dynamic}.
Consequently, each DMD mode corresponds to a distinct eigenvalue $\lambda = a + ib$, wherein $b$ denotes the oscillation frequency and $a$ represents the rate of growth or decay.

It is a natural and attractive idea to use 
the DMD method for simulating dynamics of open quantum systems.
As known,
simulating open quantum systems is a significant challenge, despite the availability various exact or approximate methods, such as the quantum jump operator method \cite{plenio1998quantum}, the quantum master equation \cite{xu2005exact,li2005quantum}, the quantum random walk method \cite{anderson1980quantum}, the quantum Monte Carlo method \cite{kolorenvc2011applications,luchow2011quantum,carlson2015quantum}, the machine learning time-local generators \cite{mazza2021machine}, and the dissipaton equation of motion (DEOM) \cite{yan2014theory,zhang2015nonperturbative,zhang2018statistical,wang2020entangled}.
The DEOM serves as the benchmark method exploited in this work, which is the second quantization generalization of the well-known hierarchical equations of motion (HEOM) \cite{tanimura1990nonperturbative,tanimura2006stochastic,yan2004hierarchical,xu2005exact,xu2007dynamics,jin2008exact,chen2016accuracy,blau2018local,dunn2019removing,ikeda2020generalization,lindoy2023quantum,ullah2021speeding,ullah2022predicting}.
Employing the linear space algebra, DEOM enables the utilization of the Nakajima-Zwanzig projection operator technique. This technique allows for a focus on the dynamics of specific subspaces and the construction of non-Markovian rate kernels  \cite{shi2003new,kelly2016generalized,zhang2016kinetic}. 
However, when dealing with complex systems, 
the huge computational cost will largely limit the applicability of these methods.
For example, the computational bottleneck emerges when calculating the the non-Markovian rate kernels in DEOM simulations, especially for large systems at low temperatures.
To overcome this shortcoming, we use the DMD algorithm
to compute snapshots of non-Markovian rate kernels within short span of time and subsequently predict the rate kernel information for the entire time window. 
This approach significantly reduces the computational resource required for simulating the quantum rate processes.

The remainder of this paper is organized as follows. Sec.~\ref{sec:theory} provides an overview of the theoretical framework of the DMD method, 
followed by the introduction of the Hamiltonian for the electron transfer system, as well as the rate kernels.
In Sec.~\ref{sec:results}, we present the numerical results for the rate kernels, population or coherence dynamics, 
exemplified with the non-Markovian population rate in both the time-independent and the Floquet scenarios.
Finally, we summarize our paper in Sec.~\ref{sec:conclusions}.

\section{Theory}
\label{sec:theory}
\subsection{Dynamic mode decomposition}
In this section, we provide a concise overview of the fundamental principles underlying the DMD technique, as well as the numerical procedures for implementing this method.

DMD is a data-driven technique to extract significant spatial modes and temporal frequencies from a nonlinear dynamical system to reduce the large number of degrees of freedom  \cite{schmid2010dynamic,kutz2016dynamic,towne2018spectral, mohan2018data}. The extracted modes and frequencies are then used to predict the future states of the nonlinear system. To be more explicit, 
let us consider a dynamical system governed by the following nonlinear ordinary differential equation:
\begin{equation}
    \frac{d\mathbf{x}(t)}{dt} = \mathbf{f}(\mathbf{x}(t),t), \quad t \ge 0.
\label{eqn:dxdt}
\end{equation}
Here, $\mathbf{x}(t):= [x_1(t),x_2(t),...,x_n(t)]^{T} \in \mathbb{C}^{n}$ is the time-dependent state variable, and $\mathbf{f}:\mathbb{C}^{n} \otimes \mathbb{R}^{+} \rightarrow \mathbb{C}^{n}$ is a nonlinear function of $\mathbf{x}$ and time $t$. The overall goal of DMD is to identify a collection of time-independent spatial modes $\phi_1,\phi_2,...,\phi_k$ alongside a set of temporal frequencies $\omega_1, \omega_2,..., \omega_k$ to approximate $\mathbf{x}(t)$:
\begin{equation}
    \mathbf{x}(t) \approx \sum_{\ell=1}^{r} \beta_{\ell}\phi_{\ell}e^{\text{i}\omega_{\ell}t},
\label{eqn:xt}
\end{equation}
where $\beta_{\ell}$ is a set of coefficients and $r$ is the rank (which is relatively small).
In practice, the trajectory $\mathbf{x}(t)$ is not known before solving Eq.~\ref{eqn:dxdt}. We now explain how to obtain the most significant values of $\phi_{i}$ and $\omega_{i}$ from a limited number of snapshots (or samples) of $\mathbf{x}(t)$ that we can solve.

To obtain the dynamic modes $\phi_{\ell}$ and their corresponding frequencies $\omega_{\ell}$, we first map the trajectories of the nonlinear dynamics to a linear system, which can be characterized easily through a spectral decomposition of the linear operator. This strategy is referred to as Koopman theory \cite{koopman1931hamiltonian,koopman1932dynamical,takeishi2017learning,bagheri2012computational,bagheri2013koopman,mezic2013analysis,rowley2009spectral}
To be more explicit, for a scalar observable $g(\mathbf{x}(t))$ within a small time interval ($\Delta t > 0$) for a dynamical system described by Eq.~\ref{eqn:dxdt} can be characterized as: 
\begin{equation}
    g(\mathbf{x}(t+\Delta t)) = \mathcal{K}_{\Delta t} g(\mathbf{x}(t))
\label{eqn:gxt}
\end{equation}
Here, $\mathcal{K}_{\Delta t}$ is a linear operator that is independent of time ($t$) and the choice of the observable function $g(\cdot)$. In general, $\mathcal{K}_{\Delta t}$ is an infinite-dimensional linear operator that has an infinite number of eigenvalues $\{\lambda\}$ and eigenfunctions $\varphi(\mathbf{x})$ satisfy $\varphi (\mathbf{x}(t+\Delta t)) = \mathcal{K}_{\Delta t} \varphi(\mathbf{x}(t)) = \lambda \varphi(\mathbf{x}(t))$. Such that we can express a set of $n$ observable functions $g_{j}(\mathbf{x}(t))$, $j=1,2,...,n$ by 
\begin{equation}
    \begin{bmatrix}
      g_{1}(\mathbf{x}(t)) \\
      g_{2}(\mathbf{x}(t)) \\
      \cdot \\
      \cdot \\
      \cdot \\
      g_{n}(\mathbf{x}(t)) \\
    \end{bmatrix}
    = 
    \begin{bmatrix}
      \nu_{1} \ \nu_{2} \cdot\cdot\cdot \nu_{r} 
    \end{bmatrix}
    \begin{bmatrix}
      \varphi_{j1}(\mathbf{x}(t)) \\
      \varphi_{j2}(\mathbf{x}(t)) \\
      \cdot \\
      \cdot \\
      \cdot \\
      \varphi_{jr}(\mathbf{x}(t)) \\
    \end{bmatrix}.
\label{eqn:vgxt}
\end{equation}
Here $\varphi_{j1}(\mathbf{x}),...,\varphi_{jr}(\mathbf{x})$ and $\lambda_{j1},...,\lambda_{jr}$
are the eigenfunctions and eigenvalues of $\mathcal{K}_{\Delta t}$ that span a subspace with $r \in  \mathbb{N}^{+}$ dimension.

In particular, if $g_{j}(\mathbf{x}(t))$ is the $j$th component of $\mathbf{x}(t)$, i.e., $g_{j}(\mathbf{x}(t))=x_{j}(t)$, we have
\begin{widetext}
\begin{equation}
    \mathbf{x}(t+\Delta t) = 
    \begin{bmatrix}
      \mathcal{K}_{\Delta t}g_{1}(\mathbf{x}(t)) \\
      \mathcal{K}_{\Delta t}g_{2}(\mathbf{x}(t)) \\
      \cdot \\
      \cdot \\
      \cdot \\
      \mathcal{K}_{\Delta t}g_{n}(\mathbf{x}(t)) \\
    \end{bmatrix}
    = 
    \begin{bmatrix}
      \lambda_{j1}\nu_{1} \ \lambda_{j2}\nu_{2} \cdot\cdot\cdot \lambda_{jr}\nu_{r} 
    \end{bmatrix}
    \begin{bmatrix}
      \varphi_{j1}(\mathbf{x}(t)) \\
      \varphi_{j2}(\mathbf{x}(t)) \\
      \cdot \\
      \cdot \\
      \cdot \\
      \varphi_{jr}(\mathbf{x}(t)) \\
    \end{bmatrix}
    = \mathbf{A}\mathbf{x}(t),
\label{eqn:gxtplusdt}
\end{equation}
\end{widetext}
Here, we have defined 
\begin{equation}
    \mathbf{A} = 
    \begin{bmatrix}
        \lambda_{j1}\nu_{1} \ \lambda_{j2}\nu_{2} \cdot\cdot\cdot \lambda_{jr}\nu_{r} 
    \end{bmatrix}
    \begin{bmatrix}
        \nu_{1} \ \nu_{2} \cdot\cdot\cdot \nu_{r} 
    \end{bmatrix}^{\dagger}
    \in \mathbb{C}^{n \times n}
\label{eqn:A}
\end{equation}
where $(\cdot)^{\dagger}$ denotes the Moore-Penrose pseudoinverse. As a result, the dynamical system of $\mathbf{x}(t)$ is completely governed by $\mathbf{A}$. Such that $\mathbf{A}$ is the main object of interest. Notice that the Koopman operator $\mathcal{K}_{\Delta t}$ is not known in advance, such that we cannot use $\mathcal{K}_{\Delta t}$ to determine $\mathbf{A}$. Below, we offer procedures to approximate $\mathbf{A}$. 

Suppose we have a sample of predetermined snapshots of $\mathbf{x}(t)$, we can approximate $\mathbf{A}$ using these snapshots. We take the uniformly distributed samples at $t_{i}=t_{1}+(i-1)\Delta t$ (where $i=1,...,m$), such that the snapshots are represented as  $\mathbf{x}_{i} = \mathbf{x}(t_{i})$. We then determine $\mathbf{A}$ by minimizing the Frobenius norm of $\mathbf{R}(\mathbf{A})$, which is defined as 
\begin{equation}
    \mathbf{R}(\mathbf{A}) = \mathbf{A}\mathbf{X}_{1} - \mathbf{X}_{2}. 
\label{eqn:RA}
\end{equation}
Here,
$\mathbf{X}_{1} = (\mathbf{x}_{1} \ \mathbf{x}_{2} \cdot\cdot\cdot \mathbf{x}_{m-1})$ and $\mathbf{X}_{2} = (\mathbf{x}_{2} \ \mathbf{x}_{3} \cdot\cdot\cdot \mathbf{x}_{m})$. 
The least squares solution to $\min_{A}\Vert \mathbf{R}(\mathbf{A}) \Vert_{F}$ (where $\Vert \cdot \Vert_{F}$ denotes the Frobenius norm) is
\begin{equation}
 \mathbf{A}=\mathbf{X}_{2}\mathbf{X}_{1}^{\dagger}. 
\label{eqn:ax2x1d}
\end{equation}
The pseudoinverse $\mathbf{X}_{1}^{\dagger}$ can be obtained from the singular value decomposition (SVD) \cite{golub2013matrix} of $\mathbf{X}_{1}, i.e.$ 
\begin{equation}
    \mathbf{X}_{1}=\mathbf{U}\Sigma\mathbf{V}^{\ast}, 
\label{eqn:x1}
\end{equation}
with $\mathbf{U} \in \mathbb{C}^{n \times n}$, $\Sigma \in \mathbb{C}^{n \times m}$, and $\mathbf{V} \in \mathbb{C}^{m \times m}$. Here we have $\mathbf{U}^{\ast}\mathbf{U}=\mathbf{I}$ and $\mathbf{V}^{\ast}\mathbf{V}=\mathbf{I}$. 

In many cases, singular values on the diagonal of $\mathbf{\Sigma}$ decay rapidly, such that the rank of $\mathbf{\Sigma}$ is small compared to the dimension of the $\mathbf{X}_{1}$, i.e. $r \ll \min\{ n,m \}$. We can then define the projection matrices: 
\begin{equation}
    \tilde{\mathbf{U}} = \mathbf{U}(:,1:r), \
    \tilde{\Sigma} = \Sigma(1:r,1:r), \ 
    \tilde{\mathbf{V}} = \mathbf{V}(:,1:r).
\label{eqn:usv}
\end{equation}
Using the projection matrices, we can map original $\mathbf{A}$ matrix into a r-rank matrix $\tilde{\mathbf{A}}$:
\begin{equation}
    \tilde{\mathbf{A}} = \tilde{\mathbf{U}}^{\ast} \mathbf{A} \tilde{\mathbf{U}} \approx \tilde{\mathbf{U}}^{\ast}\mathbf{X}_{2}\tilde{\mathbf{V}}\tilde{\Sigma}^{-1}\tilde{\mathbf{U}}^{\ast}\tilde{\mathbf{U}} = \tilde{\mathbf{U}}^{\ast}\mathbf{X}_{2}\tilde{\mathbf{V}}\tilde{\Sigma}^{-1}.
\label{eqn:Atilde}
\end{equation}
We have used the low-rank approximation in the above equation, i.e. $\mathbf{X}_{1} \approx \tilde{\mathbf{U}}\tilde{\Sigma}\tilde{\mathbf{V}}^{\ast}$.

To proceed, we solve the eigenvalue problem of the reduced matrix:
\begin{equation}
    \tilde{\mathbf{A}}\mathbf{W} = \mathbf{W}\Lambda,
\end{equation}
where $\Lambda$ is the eigenvalue matrix, 
\begin{equation}
    \Lambda = 
    \begin{bmatrix}
        \lambda_{1} &  &  \\
         & \ddots &  \\
         &  & \lambda_{r}
    \end{bmatrix}
\end{equation}
and matrix $\mathbf{W}$ is the corresponding eigenvectors. To further represent the dynamical system in the form of Eq.~\ref{eqn:xt}, we can redefine the eigenvalue matrix:  
\begin{equation}
    \mathbf{\Omega} = \frac{\ln\Lambda}{\Delta t} = 
    \begin{bmatrix}
        \text{i}\omega_{1}^\text{DMD} & & \\
         & \ddots & \\
         & & \text{i}\omega_{r}^\text{DMD}
    \end{bmatrix},
\end{equation}
where $\omega_{l}^\text{DMD} = -\text{i} \frac{\ln\lambda_{\ell}}{\Delta t}$, for $\ell = 1,...,r$. In addition, to obtain spectral modes in the original state space of $\mathbb{C}^{n}$, we perform the following transformation:
\begin{equation}
    \Phi = \mathbf{X}_{2}\tilde{\mathbf{V}}\tilde{\Sigma}^{-1}\mathbf{W}.
\end{equation}
where the columns of $\Phi$ are referred to as the DMD modes. Such that the dynamical system  $\mathbf{x}$ can be approximated by:
\begin{equation}
    \mathbf{x}(t) \approx   \Phi \exp(\mathbf{\Omega} t)\mathbf{b} = \sum_{\ell=1}^{r}\phi_{\ell}\exp(\text{i}\omega_{\ell}^\text{DMD}t)b_{\ell}.
\label{eqn:xapprox}
\end{equation}

 The amplitude vector $\mathbf{b}$ in the above equation is left to be determined. There are two approaches to calculate $\mathbf{b}$. The first approach determines $\mathbf{b}$ directly by taking the projection of the initial value $\mathbf{x}_{1}$ onto the DMD modes using the matrix $\Phi^{\dagger}$:
\begin{equation}
    \mathbf{b} = \Phi ^{\dagger} \mathbf{x}_{1},
\end{equation}
Alternatively, $\mathbf{b}$ can be computed as the least squares fit of the approximated DMD modes on the sampled trajectories. To be more explicit, we minimize the difference between the expression $\Phi\exp(\Omega t_{j})\mathbf{b}$ and the observed data $\mathbf{x}_{j}$ over a set of $m$ sampled time points.
\begin{equation}
    \mathbf{b} = \argmin_{\Tilde{\mathbf{b}} \in \mathbb{C}^{n}} \sum_{i=1}^{m} \Vert \Phi \exp(\mathbf{\Omega} t_{i})\Tilde{\mathbf{b}} - \mathbf{x}_{i} \Vert_{l^2}^2,
\end{equation}
where $\Vert \cdot \Vert_{l^2}$ denotes the standard Euclidean norm of a vector.

The description of the DMD procedure indicates that the primary computational expense arises from the SVD calculation described in Eq.~\ref{eqn:x1}, with a complexity of $O(\min{(m^2n,nm^2)})$. Note that 
DMD does not require knowledge of the underlying dynamics given by the function $\mathbf{f}(\mathbf{x}(t), t)$ in Eq.~\ref{eqn:dxdt}. Instead, DMD utilizes data from the initial time steps and predicts the future  states of the system. Additionally, DMD reduces the computational cost by projecting the $n$-dimensional space into $r$-dimensional subspace. Thus, this method is useful for analyzing nonlinear or high-dimensional dynamical systems.

\subsection{Model Hamiltonian}
We will now apply DMD to dynamics of electron transfer system in this subsection.  
Let us consider a donor-bridge-acceptor system for electron transfer, whose Hamiltonian reads \cite{su2022electron},
\begin{equation}
\begin{aligned}
    H_{\text{T}} &= h_{\text{D}}\ketbra{\text{D}}{\text{D}} + (E^{\circ} + h_{\text{A}})\ketbra{\text{A}}{\text{A}} + H_{\text{B}} \\
    &\quad
    + V(\{ \tilde{q_k} \})(\ketbra{\text{D}}{\text{A}} + \ketbra{\text{A}}{\text{D}}).
\end{aligned}
\label{eqn:HET}
\end{equation}
Here, the potential energy function $V(\{ \tilde{q_k} \})$ depends on the coordinates of the bridge, and the Hamiltonian for the fluctuating bridges can be expressed as $H_{\text{B}} = \sum_{k}\frac{\tilde{\omega}_k}{2}(\tilde{p}_{k}^2 + \tilde{q}_{k}^2)$. 
In Eq.~\ref{eqn:HET}, $E^{\circ}$ represents the standard reaction Gibbs energy for the electron transfer process from the donor state ($\ket{\text{D}}$) to the acceptor state ($\ket{\text{A}}$). 
The donor state and the acceptor state are each influenced by their respective solvent environments, $h_{\text{D}}$ and $h_{\text{A}}$. 
Specifically, $h_{\text{D}}$ can be described as $h_{\text{D}} = \sum_j\frac{\omega_j}{2}(p_j^2 + x_j^2)$, while $h_{\text{A}}$ is given by $h_{\text{A}} = \sum_j \frac{\omega_j}{2}[p_j^2 + (x_j-d_j)^2]$.
Initially, the total density operator can be represented as 
$\rho_{\text{T}}(t_0) = \ketbra{\text{D}}{\text{D}}\otimes(e^{-\beta h_{\text{D}}}/ \text{tr} e^{-\beta h_{\text{D}}}) \otimes (e^{-\beta H_{\text{B}}}/ \text{tr} e^{-\beta H_{\text{B}}}) $, which denotes the thermal equilibrium in the donor state, with $\beta=1/(k_B T)$ being the inverse temperature.

We also consider the presence of external periodic driving, which is referred to as the Floquet scenario. In such a case, we introduce periodic variation in $E^{\circ}$:
\begin{equation}\label{Fw}
    E^{\circ} \rightarrow E(t) = E^{\circ} + \varepsilon \cos{\Omega t}
\end{equation}
where $\varepsilon$ represents the driving amplitude and $\Omega$ represents the driving frequency.

The Hamiltonian in Eq.~\ref{eqn:HET} can be decomposed  into the system and the environment as follows 
\cite{su2022electron}:
\begin{equation}
\begin{aligned}
    H_{\text{T}} = H_{\text{S}} + h_{\text{E}} - \ketbra{\text{A}}{\text{A}}\delta\hat{U} - \hat{Q}\delta\hat{V}
\end{aligned}
\label{eqn:HT}
\end{equation}
Here $h_{\text{E}} = h_{\text{D}} + H_{\text{B}}$ is the Hamiltonian for the environment. $\hat{Q}$ is defined as $ \hat{Q}= \ketbra{\text{D}}{\text{A}} + \ketbra{\text{A}}{\text{D}}$. $H_{\text{S}}$ is the system Hamiltonian given by 
\begin{equation}
\begin{aligned}
    H_{\text{S}} = &(E^{\circ} + \varepsilon \cos{\Omega t} + \lambda)\ketbra{\text{A}}{\text{A}} 
    \\ \quad
    &+ \expval{V}_{\text{B}}
    (\ketbra{\text{D}}{\text{A}} + \ketbra{\text{A}}{\text{D}}).
\end{aligned}
\label{eqn:HS}
\end{equation}
In  Eqs.~\ref{eqn:HT} and \ref{eqn:HS}, $\delta \hat U\equiv\hat U-\langle\hat U \rangle_{\text{D}}$ with 
$\hat U\equiv h_{\text{A}}-h_{\text{D}}$ and $\langle\hat U \rangle_{\text{D}}\equiv \text{tr}(\hat U e^{-\beta h_{\text{D}}})/ \text{tr} (e^{-\beta h_{\text{D}}})$,
whereas $\delta \hat V\equiv
\langle V\rangle_{\text{B}}-V(\{ \tilde{q_k} \})=\sum_{k}\tilde{c_k}\tilde{q}_k$ with $\langle V \rangle_{\text{B}}\equiv\text{tr}(V e^{-\beta H_{\text{B}}})/ \text{tr} (e^{-\beta H_{\text{B}}})$.

In the simulations, we incorporate the spectral densities, $J_{\text{D}}(\omega) \equiv (1/2)\int_{-\infty}^{\infty} {\rm d}t e^{i\omega t} \langle [\delta \hat{U}(t), \delta \hat{U}(0)] \rangle_{\text{D}}$ and $J_{\text{B}}(\omega) \equiv (1/2)\int_{-\infty}^{\infty} {\rm d}t e^{i\omega t} \langle [\delta \hat{V}(t), \delta \hat{V}(0)] \rangle_{\text{B}}$, as \cite{weiss2012quantum,kleinert2009path,yan2005quantum}
\begin{equation}
    J_{\text{D}}(\omega) = \frac{2\lambda \gamma \omega}{\omega^2 + \gamma^2}
\label{eqn:JD}
\end{equation}
and 
\begin{equation}
    J_{\text{B}}(\omega) = \frac{2\lambda' 
    \omega_0^2 \zeta \omega}{(\omega^2 - \omega_{0}^2)^2 + \omega^2 \zeta^2},
\label{eqn:JB}
\end{equation}
respectively.

\subsection{Rate kernel calculation}
In this subsection, we present the DEOM of the rate kernels. 
From a theoretical perspective, it is possible to precisely construct the generalized rate equation as 
\begin{equation}
\begin{aligned}
    \dot{P}_\text{D}(t) =& - \int _{0}^{t} d\tau k(t-\tau;t)P_{\text{D}}(\tau) \\
    &+ \int _{0}^{t} d\tau k'(t-\tau;t)P_{\text{A}}(\tau).
\end{aligned}
\label{eqn:PDt}
\end{equation}
Here, $P_{\text{D}}(t)$ and $P_{\text{A}}(t)$ denote the populations of the donor and acceptor, respectively. Here, we employ the forward and backward rate memory kernels ($k(\tau;t)$ and $k'(\tau;t)$) in the rate equation. $\tau$ is the memory time scale, capturing the non-Markovian nature of the system. Note that $t$ in $k(\tau;t)$ and $k'(\tau;t)$) represents the time dependence of the rate kernels due to the presence of an external field. These rate kernels are regularly calculated and examined to investigate the memory effect in rate processes\cite{gong2015continued, zhang2016kinetic, xu2018convergence, yan2019theoretical, dan2022generalized}.

The rate kernels are constructed using the generalized master equation \cite{zhang2016kinetic}. Based on the composite Hamiltonian in Eq.~\ref{eqn:HET}, the DEOM can be expressed as
\begin{equation}
    \dot{\bm{\mathbf{\rho}}}(t) = -i\bm{\mathcal{L}}(t)\bm{\rho}(t).
\label{eqn:deom}
\end{equation}
This equation is similar to the time evolution equation for the total system, $\dot{\rho}_\text{T} = -i\mathcal{L}_\text{T}(t)\rho_\text{T}$, despite that the total Liouvillian $\mathcal{L}_\text{T}(t)$ is mapped to the DEOM-space dynamics generator $\bm{\mathcal{L}}(t)$, and the total density matrix $\rho_\text{T}(t)$ is mapped to the DEOM density matrix $\bm{\rho}(t) = \{ \rho_{\bf n}^{(n)}(t);n=0,1,2,... \}$. Here, $\mathcal{L}_{\text{T}}(t) \equiv [H_{\text{T}}(t), \cdot]$. To proceed, we define the projection operators in DEOM space, $\bm{\mathcal{P}}$ and $\bm{\mathcal{Q}} = \bm{\mathcal{I}} - \bm{\mathcal{P}}$, to separate $\bm{\rho}$ into its population and coherence components \cite{zhang2016kinetic}:
\begin{equation}
\begin{aligned}
  \bm{\mathcal{P}} \bm{\mathbf{\rho}}(t) = \left\{ \sum_{a}\rho_{aa}^{(0)}(t)\ketbra{a}{a};\ 0,\ 0,\ \cdot\cdot\cdot\right\} \equiv \bm{p}(t)  \\
  \bm{\mathcal{Q}} \bm{\mathbf{\rho}}(t) = \left\{ \sum_{a \neq b}\rho_{ab}^{(0)}(t)\ketbra{a}{b};\rho_{\bf n}^{(n>0)}(t)\right\} \equiv \bm{\mathbf{\sigma}}(t).
\end{aligned}
\label{eqn:ProjectPQ}
\end{equation}
We can now express the DEOM in Eq.~\ref{eqn:deom} in a different form by using the following matrix representation: 
\begin{equation}
\begin{aligned}
    \begin{bmatrix}
        \dot{\bm{p}}(t)\\
        \dot{\bm{\mathbf{\sigma}}}(t)
    \end{bmatrix}
    = -i
    \begin{bmatrix}
        \bm{\mathcal{P}} \bm{\mathcal{L}}(t) \bm{\mathcal{P}} & \bm{\mathcal{P}} \bm{\mathcal{L}}(t) \bm{\mathcal{Q}} \\
        \bm{\mathcal{Q}} \bm{\mathcal{L}}(t) \bm{\mathcal{P}} & \bm{\mathcal{Q}} \bm{\mathcal{L}}(t) \bm{\mathcal{Q}} 
    \end{bmatrix}
    \begin{bmatrix}
        \bm{p}(t)\\
        \bm{\mathbf{\sigma}}(t)
    \end{bmatrix}.
\end{aligned}
\label{eqn:newdeom}
\end{equation}
Similar to the procedure in Nakajima–Zwanzig equation \cite{nakajima1958quantum,zwanzig1960ensemble,zhang2016kinetic}, we can obtain the equation of motion for $\bm{p}(t)$ as follows:
\begin{equation}
\begin{aligned}
    \dot{\bm{p}}(t) = \int_0^t d\tau \Tilde{\bm{K}}(t-\tau;t) \bm{p}(\tau),
\end{aligned}
\label{eqn:pdot}
\end{equation}
where the rate kernel $\Tilde{\bm{K}}(t-\tau;t)$ can be expressed as:
\begin{equation}
\begin{aligned}
    \Tilde{\bm{K}}(t-\tau;t)=-\bm{\mathcal{P}} \bm{\mathcal{L}}(t) \bm{\mathcal{Q}} \bm{\mathcal{U}}(t,\tau) \bm{\mathcal{Q}} \bm{\mathcal{L}}(\tau) \bm{\mathcal{P}}.
\end{aligned}
\label{eqn:ratekernel}
\end{equation}
In Eq.~\ref{eqn:Uttau}, we define the time-evolution operator $\bm{\mathcal{U}}(t,\tau)$ as:
\begin{equation}
\begin{aligned}
 \bm{\mathcal{U}}(t,\tau) \equiv \exp_{+}{\left [-i \int_\tau^t d\tau'\bm{\mathcal{L}}(\tau') \right ]}.
\end{aligned}
\label{eqn:Uttau}
\end{equation}
It is worth noting that $-k(t-\tau;t)$ and $k(t-\tau;t)$ in Eq.~\ref{eqn:PDt} correspond to the $\ketbra{\text{D}}{\text{D}} \rightarrow \ketbra{\text{D}}{\text{D}}$ and $\ketbra{\text{A}}{\text{A}} \rightarrow \ketbra{\text{D}}{\text{D}}$ components of $\Tilde{\bm{K}}(t-\tau;t)$, respectively.

\section{Results and discussions}
\label{sec:results}
\subsection{Population rate without Floquet driving}
\label{time_independent}
\begin{center}
\begin{figure}[h] 
 \includegraphics[width=.48\textwidth]{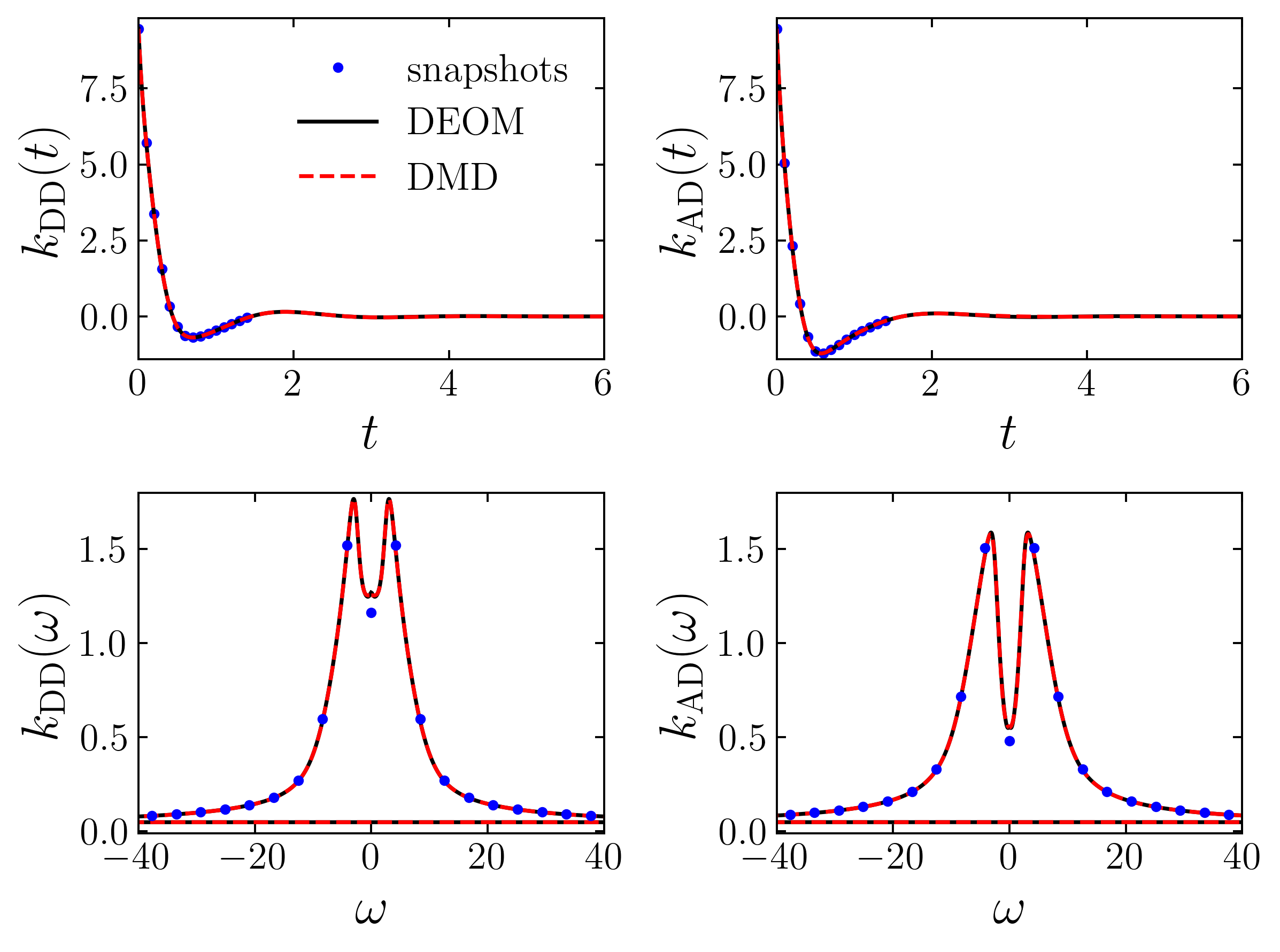}
 \caption{The rate kernels $k_{\text{DD/AD}}(t)$ (upper panel) and their Fourier transformations $k_{\text{DD/AD}}(\omega)$ (lower panel), obtained through three methods: snapshots, DEOM, and DMD. $k_{\text{DD}}(t)$ and $k_{\text{AD}}(t)$ represent the rate kernels from the donor to the donor and the acceptor to the donor, respectively. Snapshots are obtained by sampling DEOM from $0 \leq t \leq 1.5$. This study utilizes a total of 150 snapshots. To ensure graph clarity, the snapshots are plotted every 10 intervals. Based on the snapshots, DMD predicts kernels of the same length as DEOM. In reality, DMD has the capability to predict for any future time. However, for visualization purposes, the results are plotted only up to $t=6$. The spectral density is represented by $J_{\text{D}}(\omega)$. $H_{\text{S}}$ is defined as $\sigma_x + \sigma_z$, where $\sigma$ refers to the Pauli matrices. The time step is $dt=0.01$. Additionally, we have $\lambda=1$, $\gamma=\omega_0=\zeta=1$.}
\label{fig:k}
\end{figure}    
\end{center}

If the system is time-independent, the forward and backward rate memory kernels [$k(t-\tau;t)$ and $k'(t-\tau;t)$)] in Eq.~\ref{eqn:PDt} can be simplified  as $k_{\text{DD}}(t-\tau)$ and $k_{\text{AD}}(t-\tau)$, respectively, 
due to the time--translation invariance.
We first obtain a small number of snapshots of the rate kernels calculated by DEOM.
Then the DMD is used to predict the remaining part based on these snapshots, as depicted in upper panel of Fig.~\ref{fig:k}. 

Without an external field, we observe that both rate kernels $k_{\text{DD}}(t)$ and $k_{\text{AD}}(t)$ decrease to a value that slightly below zero, followed by an increase to a value slightly above zero, until ultimately approaching to zero.
We see that DMD can accurately predict the rate kernels for short and long time, which agree with DEOM results completely.
In lower panel of Fig.~\ref{fig:k}, we plot the Fourier transformation of the rate kernels. Notice that  
the rate kernels predicted by the DMD is more accurate than 
that obtained directly from the snapshots, especially near $\omega = 0$. This is due to the fact that DMD provides accurate dynamics in longer time, which results in precise spectra near $\omega = 0$. 

\begin{figure}[htbp]
\includegraphics[width=.48\textwidth]{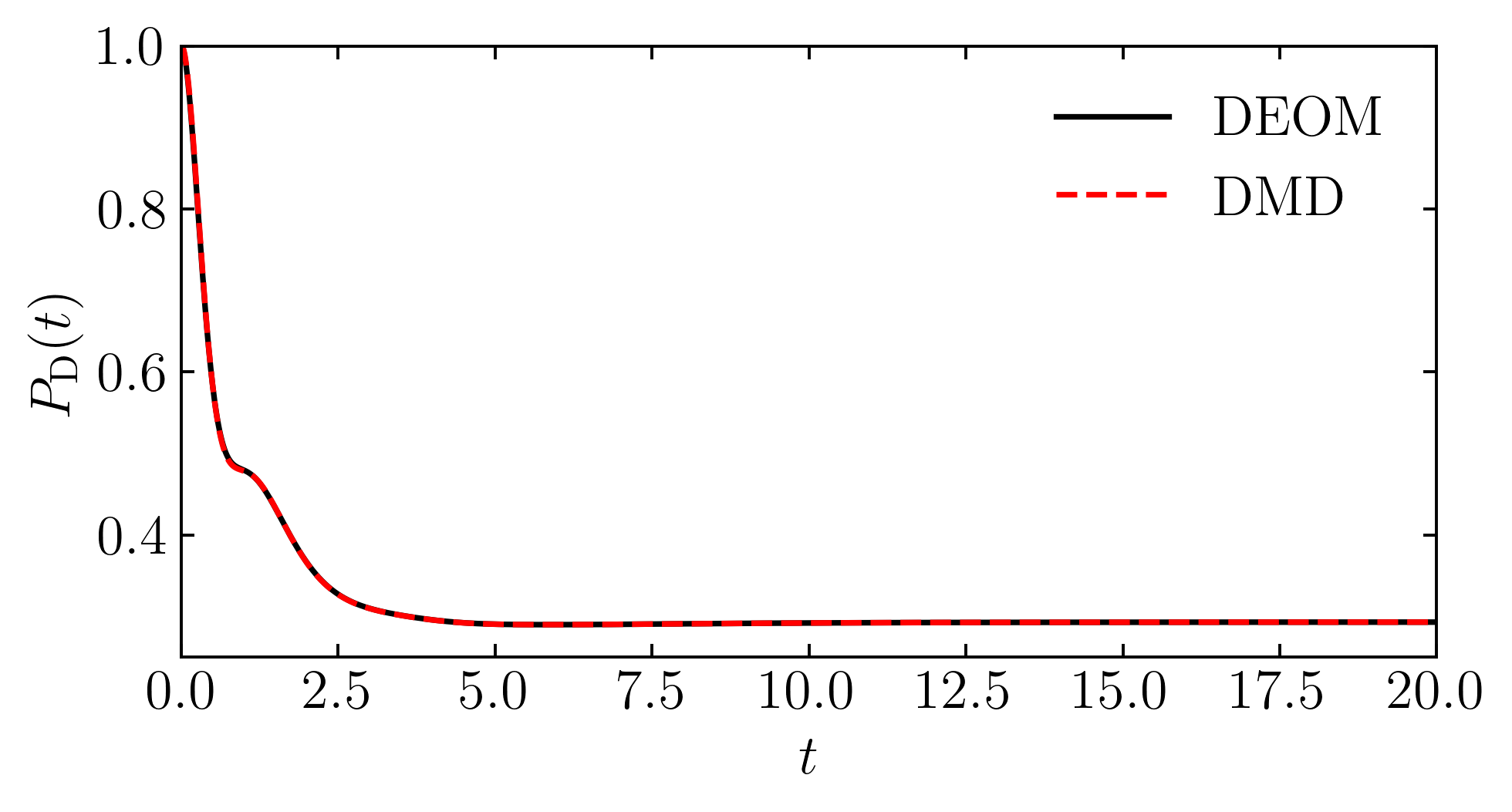}
\caption{The population of the donor, $P_{\text{D}}(t)$, is computed using the kernels depicted in Fig.~\ref{fig:k} and Eq.~\ref{eqn:PDt}. All the parameters used are identical to those in Fig.~\ref{fig:k}.
} 
\label{fig5}
\end{figure}

As a validation, we use the rate kernels $k_{\text{DD}}(t)$ and $k_{\text{AD}}(t)$ obtained through  the DMD to calculate donor population $P_{\text{D}}(t)$ by Eq.~\ref{eqn:PDt}. We also plot $P_{\text{D}}(t)$ obtained directly from DEOM calculation as a benchmark in Fig.~\ref{fig5}. 
In the DMD calculation, we have used $150$ snapshots with $dt=0.01$. We can see that DMD results agree with DEOM results perfectly. 
The DMD accuracy can be further enhanced if the number of snapshots are increased and the value of $dt$ is decreased.
Importantly, note that only a small number of snapshots for a short time of the rate kernels are required in DMD, such that DMD  significantly reduces the overall computation cost. 
Additionally, DMD is capable of predicting observables for any future time.


\subsection{Population rate with Floquet driving}
\label{floquet_system}
If the system is Floquet-driven, the forward and backward rate memory kernels in Eq.~\ref{eqn:PDt} will 
satisfy
\begin{align}
 k(t-\tau;t+T_{0})=k(t-\tau;t),
 \\
  k'(t-\tau;t+T_{0})=k'(t-\tau;t).
\end{align}
 Here, $T_{0}=2\pi/\Omega$ is the period of the Floquet driving [cf.\,Eq.~\ref{Fw}].
 We use $k^{\text{DD}}$ and $k^{\text{AD}}$ to denote the rate kernel
 from donor to donor $k$  and acceptor to donor $k'$ , respectively.
Obviously, $k^{\text{DD}}(\tau;t)$ and $k^{\text{AD}}(\tau;t)$ have two independent variables, $\tau$ and $t$. Due to the periodicity of the rate kernels, we perform a Fourier expansion on $t$,
\begin{equation}
    k^{\text{DD/AD}}(\tau;t) = \sum_{n=-\infty}^{\infty} k_n^{\text{DD/AD}}(\tau)e^{\text{i}n\Omega t}
\label{eqn:ktaut}
\end{equation}
Here the Fourier component $\{ k_n^{\text{DD/AD}}(\tau) \}$ is given by 
\begin{equation}
    k_n^{\text{DD/AD}}(\tau) = \frac{1}{T_0}\int_{0}^{T_0} dt k^{\text{DD/AD}}(\tau;t)e^{\text{i}n\Omega t}.
\end{equation}
Since the Fourier component approaches zero rapidly with increasing $n$, we only need to compute finite number of the Fourier components. Also, due to the symmetry $k_n^{\text{DD/AD}}(\tau)=k_{-n}^{\text{DD/AD}}(\tau)^{*}$, we only need to calculate Fourier components with $n \geq 0$.

\begin{center}
\begin{figure}[h] 
 \includegraphics[width=.48\textwidth]{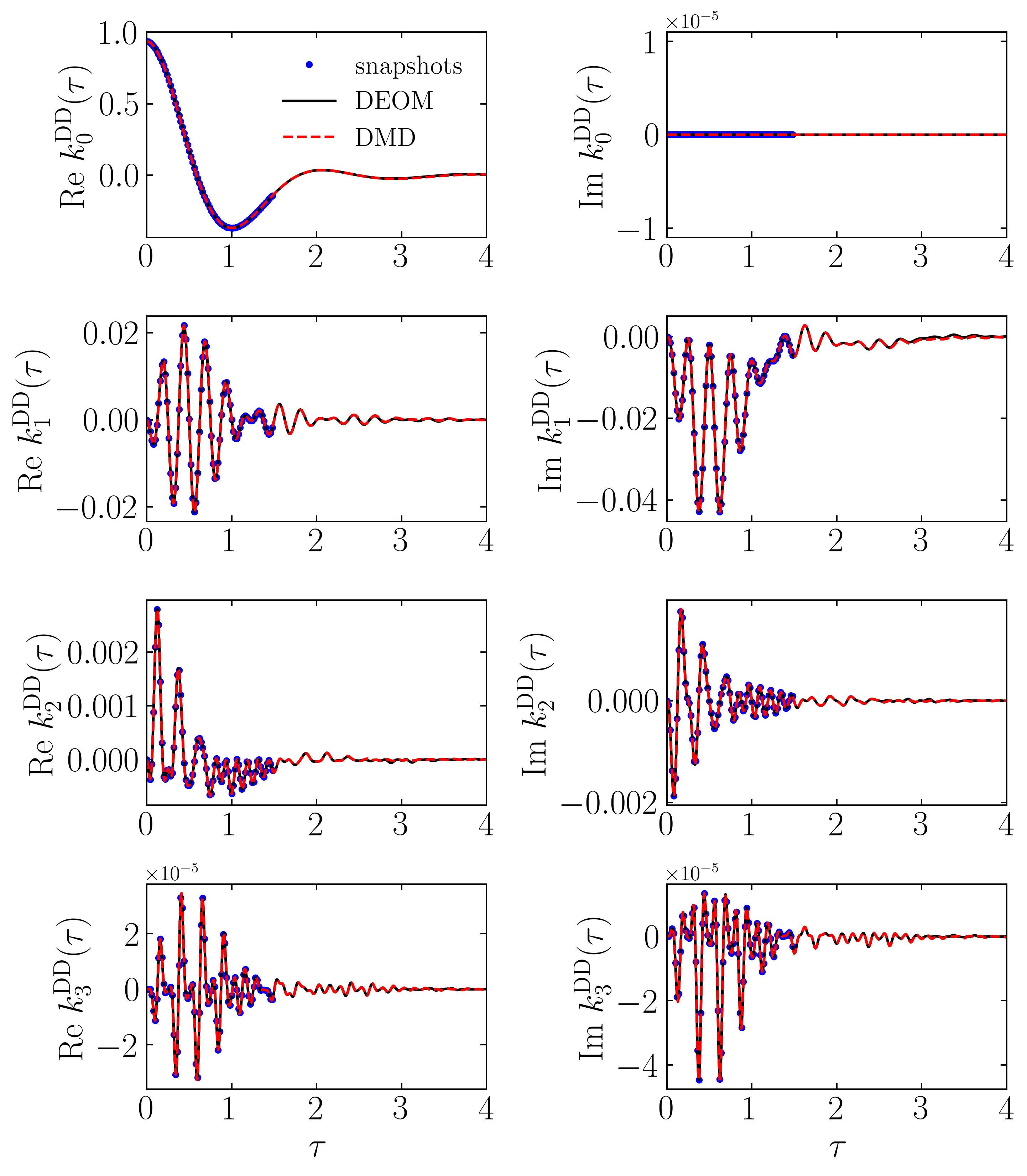}
 \caption{The Fourier expansion terms of the kernel from donor to donor $k_n^{\text{DD}}$ and $n=0,1,2,3$, obtained by three methods: snapshots, DEOM and DMD. Among them, snapshots are obtained by sampling of DEOM, and DMD predicts kernels of the same length as DEOM based on the snapshots. The spectral densities are $J_{\text{D}}(\omega)$ and $J_{\text{B}}(\omega)$, $dt=0.01$, and the number of snapshots is 150. Besides, $\varepsilon=2, \Omega=4, E^{\circ}=1.5, \lambda=\lambda'=0.2$ and $\gamma=\omega_0 =\zeta=1$.
 }
\label{fig:kn_DD}
\end{figure}    
\end{center}

\begin{center}
\begin{figure}[h] 
 \includegraphics[width=.48\textwidth]{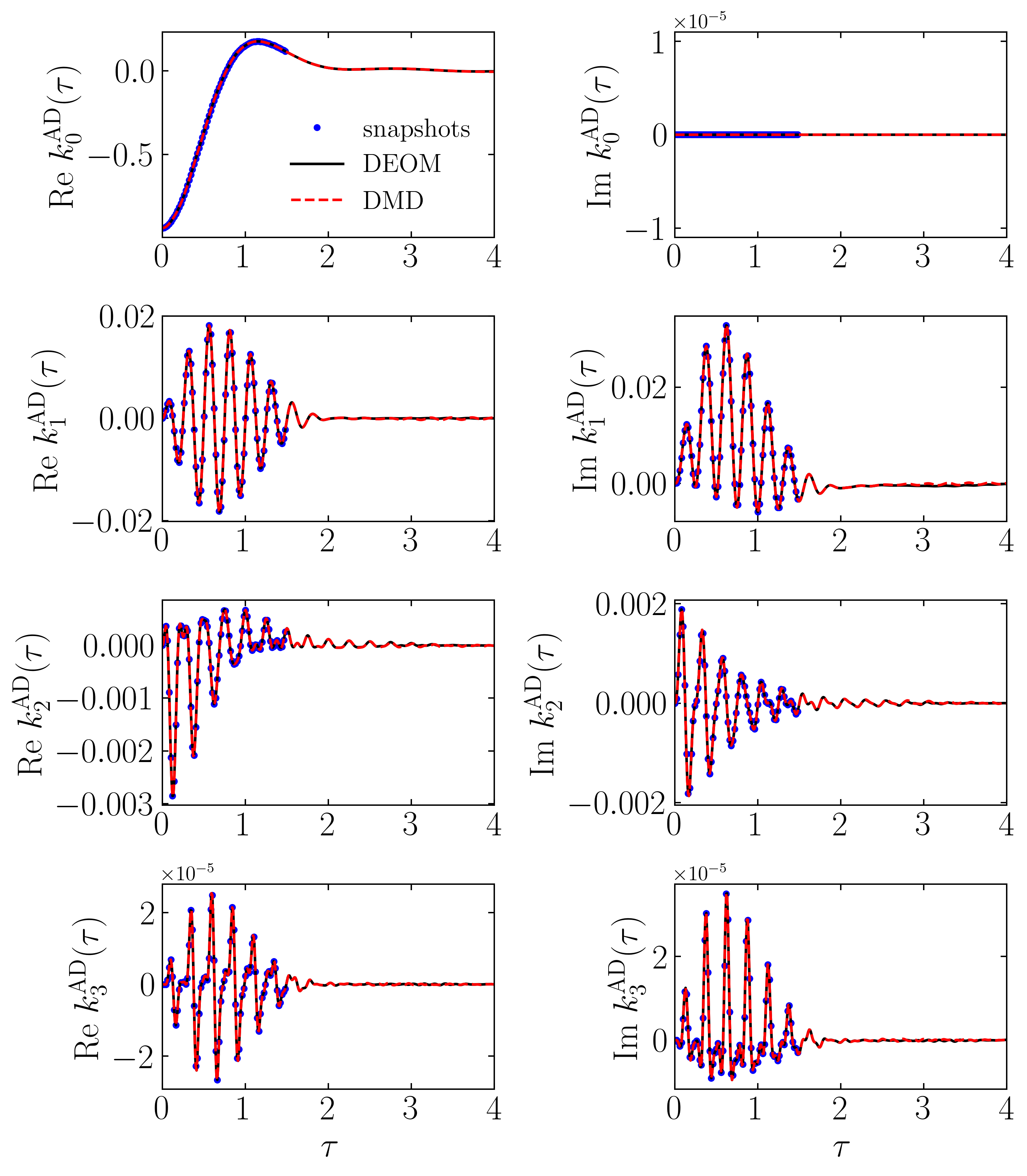}
 \caption{The Fourier expansion terms of the kernel from acceptor to donor $k_n^{\text{AD}}$ and $n=0,1,2,3$, obtained by three methods: snapshots, DEOM and DMD, with the same parameters used in Fig.~\ref{fig:kn_DD}.}
\label{fig:kn_AD}
\end{figure}    
\end{center}

We use $J_{\text{D}}(\omega)$ in Eq.~\ref{eqn:JD} and $J_{\text{B}}(\omega)$ in Eq.~\ref{eqn:JB} as spectral densities in the DEOM calculation and the initial state is on the donor. We utilize DMD to predict the Fourier components $k_n^{\text{DD}}(\tau)$ and $k_n^{\text{AD}}(\tau)$ based on the snapshots sampled from DEOM results as shown in Fig.~\ref{fig:kn_DD} and Fig.~\ref{fig:kn_AD}, respectively.
As we can see, unlike in the time-independent 
scenario,
in the case where $n$ is non-zero, the Fourier expansion terms of the rate kernels exhibit oscillation in both the real and imaginary parts. 
Furthermore, with increasing $n$, the magnitudes of the Fourier components  decrease rapidly, ultimately converging to zero. Notice that even we do not sample $k_n(\tau)$ until convergence to zero, DMD can accurately predict long time behavior. This is due to the fact that the temporal frequencies and spatial modes can be extracted efficiently from the short time dynamics, such that DMD can predict future results with sufficiently small error. 
In our example, we have used only the snapshots within the range $0 \leq t \leq 1.5$ with $dt=0.01$. Still, the predicted results obtained using DMD are nearly identical to those obtained using DEOM. This greatly reduces the overall amount of computation time.

\begin{figure}[htbp]
\includegraphics[width=.48\textwidth]{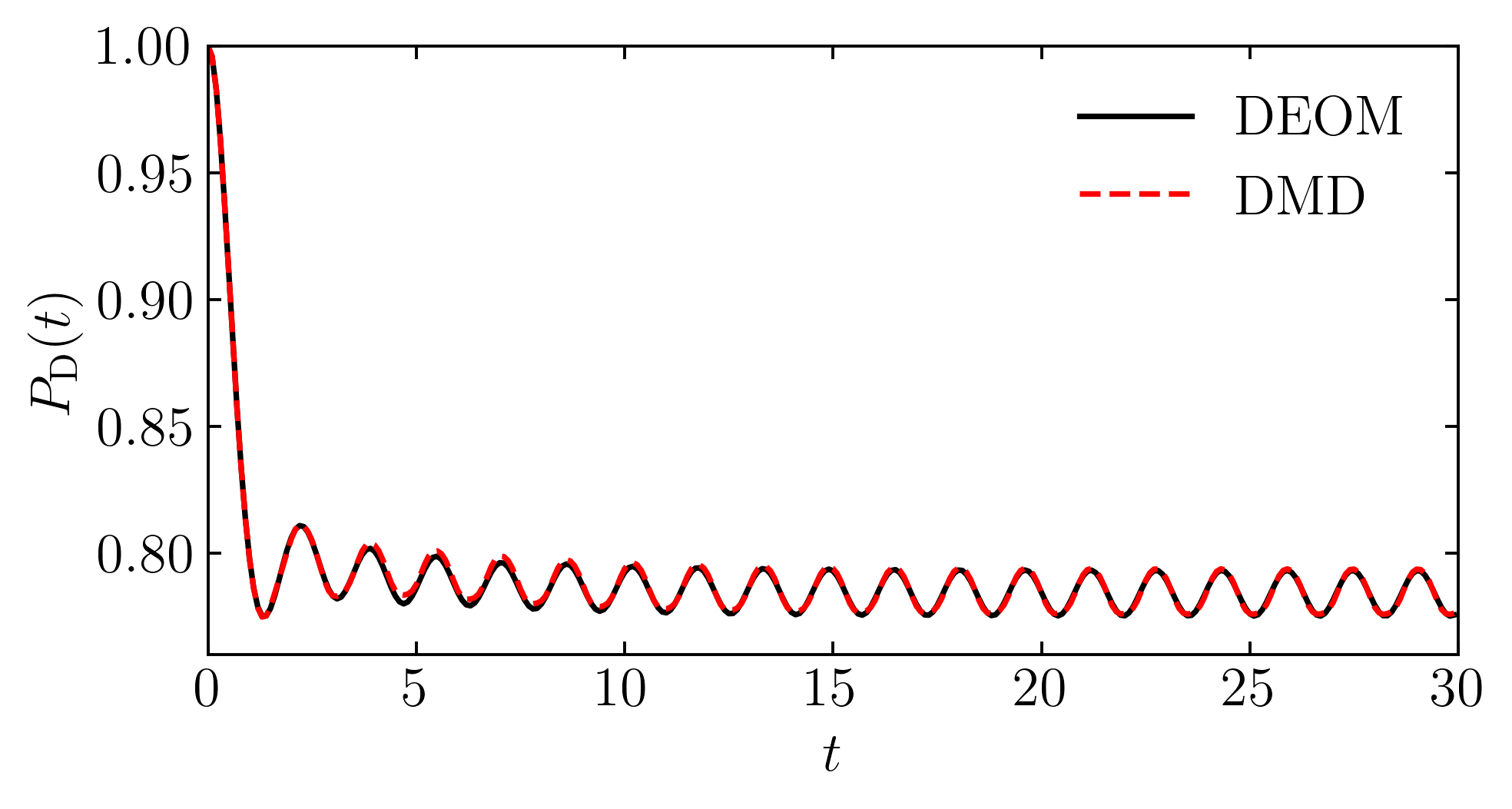}
\caption{The population of the donor $P_{\text{D}}(t)$ calculated using the kernels shown in Fig.~\ref{fig:kn_DD} and Fig.~\ref{fig:kn_AD} together with Eq.~\ref{eqn:ktaut} and Eq.~\ref{eqn:PDt}. All the parameters are the same as those used in Fig.~\ref{fig:kn_DD}.}
\label{fig:f_PDt}
\end{figure}

To further verify our prediction, we substitute the values of $k_n^{\text{DD}/\text{AD}}(\tau)$ obtained using DMD in Fig.~\ref{fig:kn_DD} and Fig.~\ref{fig:kn_AD} back into the rate kernels (in Eq.~\ref{eqn:ktaut}) and solve for donor population $P_\text{D}(t)$ (using Eq.~\ref{eqn:PDt}). The results are illustrated in Fig.~\ref{fig:f_PDt}. Under the influence of Floquet driving, $P_\text{D}(t)$ reaches to a limit cycle instead of a steady state. The period of the limit cycle is equal to the period of the driving frequency. This proves that DMD can even predict long time behavior that is not a steady state.  The accuracy of predicting the period of this oscillation mainly depends on the accuracy of DMD in predicting the imaginary part of $k_n^{\text{DD}/\text{AD}}(\tau)$, which jointly determines the range of ultimate convergence along with the real part of $k_n^{\text{DD}/\text{AD}}(\tau)$.

\subsection{Density matrix evolution}
In both Sec.~\ref{time_independent} and Sec.~\ref{floquet_system}, we obtain $P_\text{D}(t)$ by integrating only the rate kennels.
In this subsection, we utilize DMD to predict the complete kernel tensor $\bm{\mathcal{K}}(t)$ and evolve the density matrix $\bm{\rho}(t)$ to obtain $P_\text{D}(t)$ and coherence evolution simultaneously.  
In analogy to  Eq.~\ref{eqn:pdot}, the equation of motion for the density operator reads

\begin{equation}
    \dot{\bm{\rho}}(t) = -\text{i}[H_s, \bm{\rho}(t)] + \int_0^t d\tau \bm{\mathcal{K}}(t-\tau) \bm{\rho}(t)
\label{eqn:rho}
\end{equation}
where
\begin{equation}
    \bm{\rho}(t) = 
    \begin{bmatrix}
        \rho_{\text{DD}}(t) & \rho_{\text{DA}}(t)\\
        \rho_{\text{AD}}(t) & \rho_{\text{AA}}(t)
    \end{bmatrix},
\end{equation}
and
\begin{equation}
    \bm{\mathcal{K}}(t) = 
    \begin{bmatrix}
        \bm{K}_{\text{DD}}(t) & \bm{K}_{\text{DA}}(t)\\
        \bm{K}_{\text{AD}}(t) & \bm{K}_{\text{AA}}(t)
    \end{bmatrix}.
\end{equation}
Here, $\rho_{ij}(t)$ satisfies $\rho_{\text{DD}} + \rho_{\text{AA}} = 1$ and $\rho_{\text{DA}} = \rho^{\ast}_{\text{AD}}$. 
As a result,
$\bm{K}_{\text{DD}}(t)=-\bm{K}_{\text{AA}}(t)$ and $\bm{K}_{\text{DA}}(t)=[\bm{K}_{\text{DA}}(t)]^{\ast}$. Therefore, we only need to calculate $\bm{K}_{\text{DD}}(t)$ and $\bm{K}_{\text{DA}}(t)$ in our simulation.

\begin{center}
\begin{figure}[h] 
 \includegraphics[width=.48\textwidth]{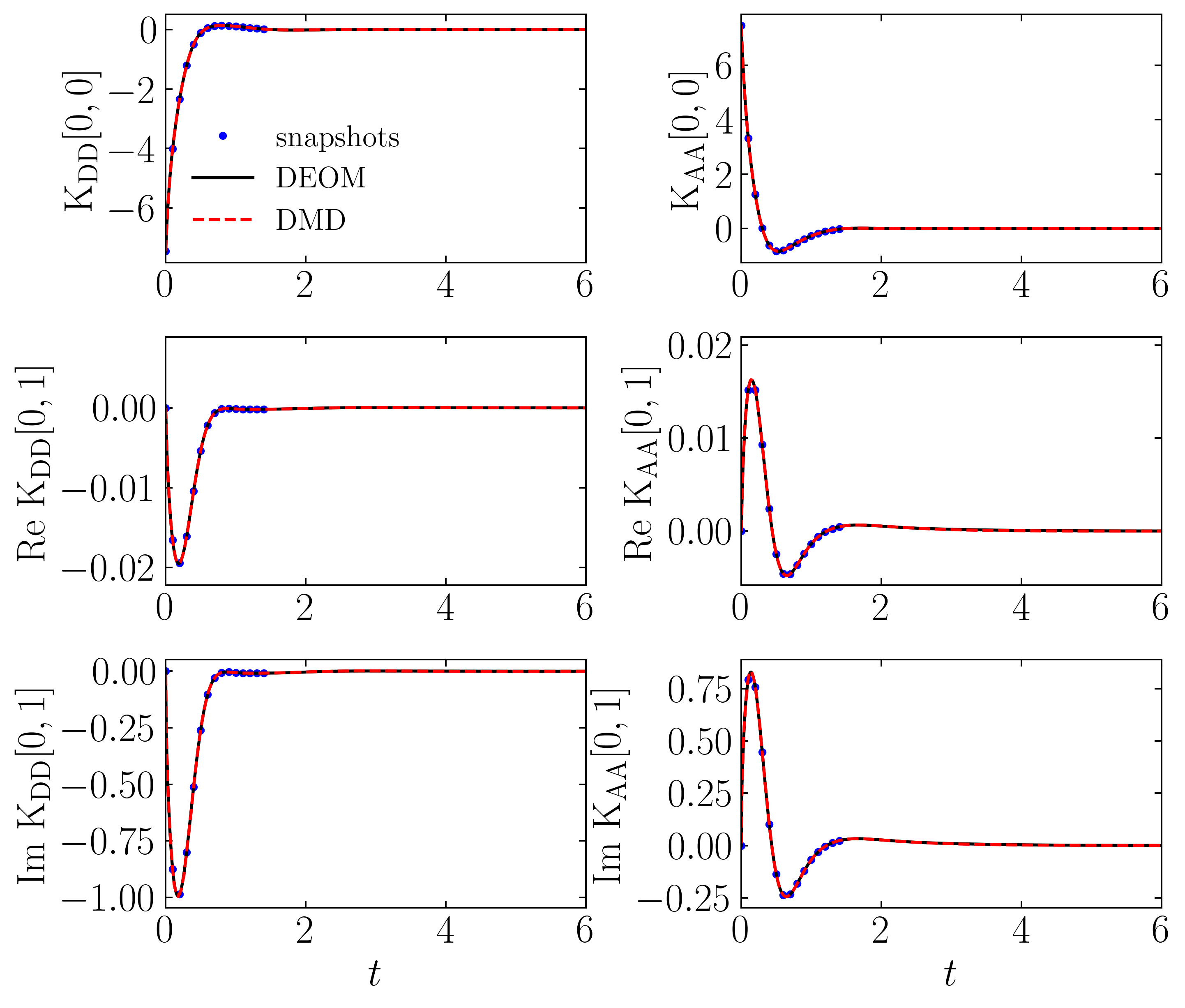}
 \caption{The elements of kernel tensors for $\bm{K}_{\text{DD}}(t)$ (left column) and $\bm{K}_{\text{AA}}(t)$ (right column) obtained by three methods: snapshots, DEOM and DMD. $\bm{K}_{\text{DD/AA}}[i,j]$ represents the i-th row and j-th column position of $\bm{K}_{\text{DD/AA}}$ at every timestep. All the parameters are the same as those used in Fig.~\ref{fig:k}.
 }
\label{fig:krhoDD_AA}
\end{figure}    
\end{center}

\begin{center}
\begin{figure}[h] 
  \includegraphics[width=.48\textwidth]{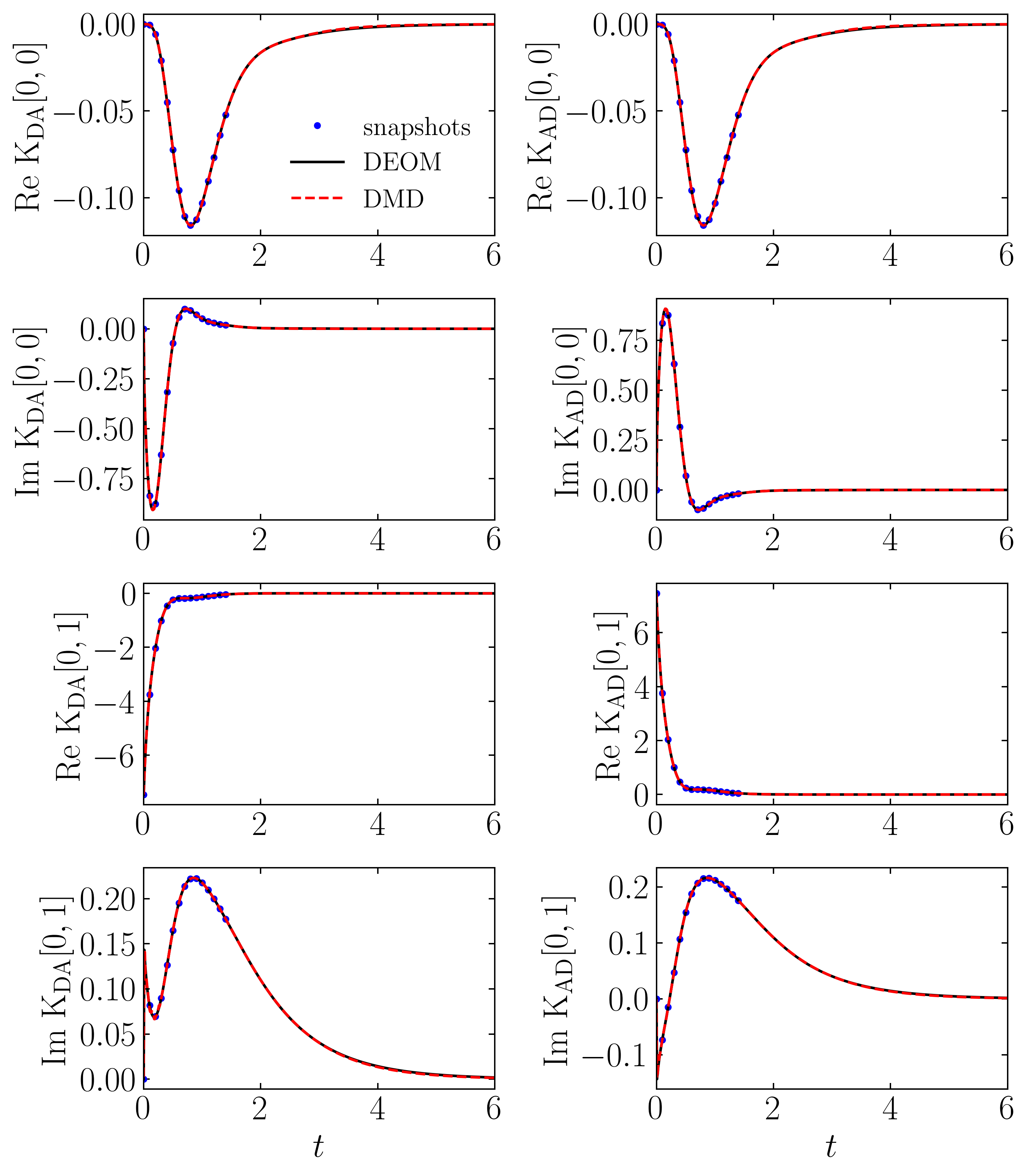}
 \caption{The elements of kernel tensor for $\bm{K}_{\text{DA}}(t)$ (left column) and $\bm{K}_{\text{AD}}(t)$ (right column) obtained by three methods: snapshots, DEOM and DMD, with the same parameters used in Fig.~\ref{fig:k}.
 }
\label{fig:krhoDA_AD}
\end{figure}    
\end{center}

\begin{center}
\begin{figure}[h] 
 \includegraphics[width=.48\textwidth]{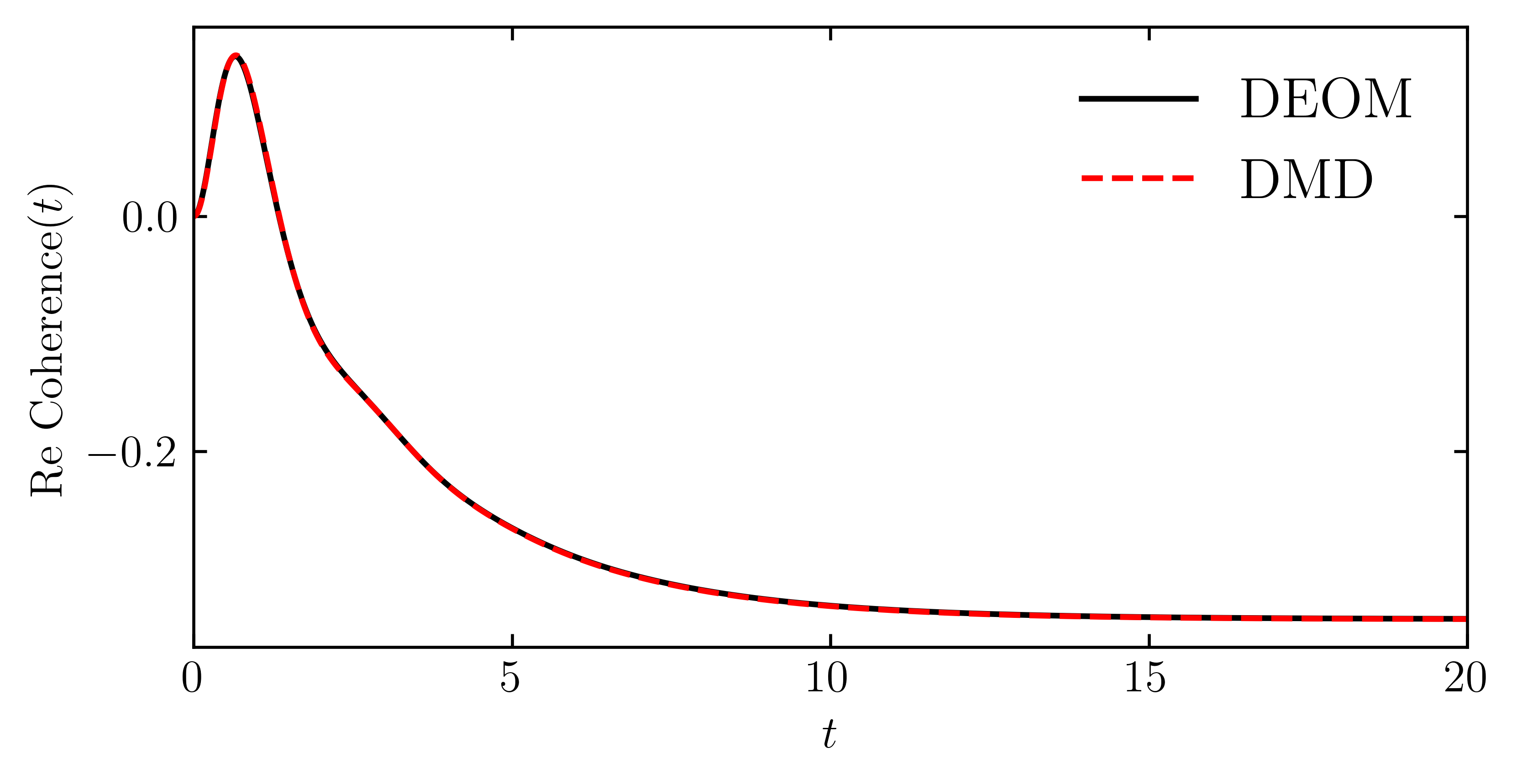}
 \includegraphics[width=.48\textwidth]{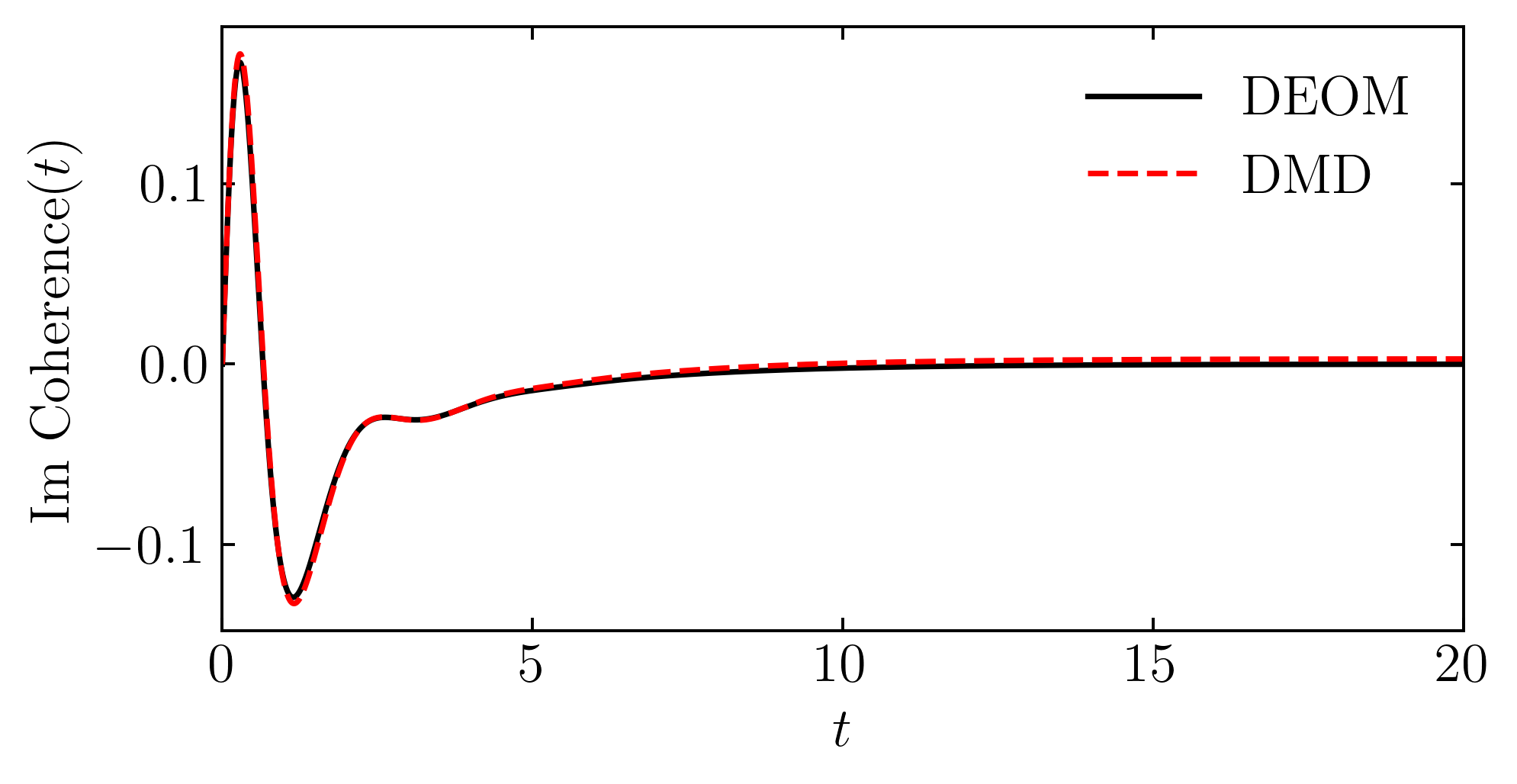}
 \caption{The coherence between donor and acceptor obtained by DEOM and DMD using the kernel tensors in Fig.~\ref{fig:krhoDD_AA} and Fig.~\ref{fig:krhoDA_AD} together with Eq.~\ref{eqn:rho}, with the same parameters used in Fig.~\ref{fig:k}}
\label{fig:dm_coherence}
\end{figure}    
\end{center}

\begin{figure}[htbp]
\includegraphics[width=.48\textwidth]{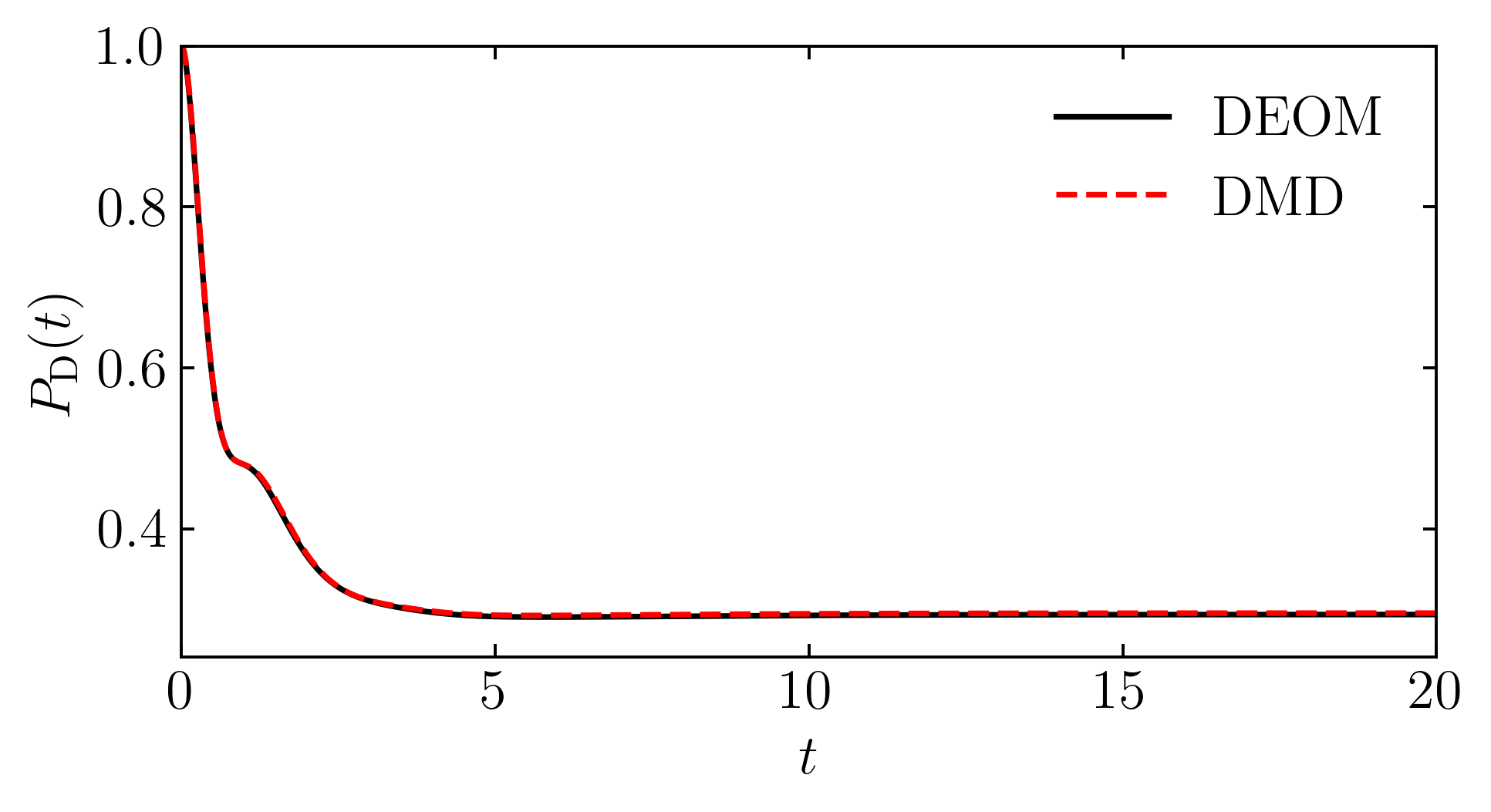}
\caption{The population of the donor $P_{\text{D}}(t)$ obtained by DEOM and DMD, with the same parameters used in Fig.~\ref{fig:k}.}
\label{fig:dm_PDt}
\end{figure}

As a example, we use the same time-independent system as in Sec.~\ref{time_independent}.
%
We obtain $150$ snapshots of the kernel tensors with $dt=0.01$, calculated by DEOM using the spectral density $J_{\text{D}}(\omega)$ in Eq.~\ref{eqn:JD} and the initial state is on the donor.
We predict the kernels using DMD based on these snapshots as shown in Fig.~\ref{fig:krhoDD_AA} and Fig.~\ref{fig:krhoDA_AD}. For all kernel tensors $\bm{K}_{ij}(t)$ , we see the DMD prediction agrees well with the DEOM calculation for both the real part and the imaginary part.
We then assemble these kernel tensors into the generalized master equation (Eq.~\ref{eqn:rho}) to calculate the time evolution of the density matrix. 
We plot the coherence $\rho_{\text{DA}}(t)$ in  Fig.~\ref{fig:dm_coherence} and the donor population $\rho_{\text{DD}}(t)$ in Fig.~\ref{fig:dm_PDt}. Note that the coherence is complex and DMD predicts both the real part and the imaginary part very well. In particular, the time-scale required for the imaginary part of the coherence approaching zero is the so-called decoherence time, which is captured well by DMD. Note that we have used the same parameters in Fig.~\ref{fig:dm_PDt} and Fig.~\ref{fig5}, and the results are identical. Different from the generalized master equation, the rate kernels do not require the information of coherence. Nevertheless, the rate kernel methods give identical results as the generalized master equation.


\section{Conclusions}
\label{sec:conclusions}
To summarize, in this work,
we exploit the DMD method to investigate the rate kernels 
in the simulation of open quantum systems.
Traditional numerical methods to obtain the rate kernels involve solving coupled two-time nonlinear integral differential equations, which results in high memory requirement and large computational cost. In contrast, the data-driven DMD method only depends on a small sampled set of the numerical solutions, and can be easily applied through the truncated SVD decomposition.
Our numerical results on the rate kernels of open quantum systems show that the DMD successfully captures the major dynamical modes and the frequencies of rate kernels, whether the external field is involved or not. %
We anticipate that the DMD would become a useful tool for simulating dynamics of open quantum systems.

\begin{acknowledgments}
W.D. acknowledges start-up funding from Westlake University.
\end{acknowledgments}

\bibliography{ref}
\end{document}